\documentclass[review]{elsarticle}

\usepackage{amsmath,amssymb}
\usepackage{lipsum,soul}
\usepackage{bm}
\usepackage{graphicx}
\usepackage{graphics}
\usepackage{amsmath}
\usepackage{array}
\usepackage[caption=false]{subfig}
\usepackage{xcolor}
\usepackage{subfig}
\usepackage{hyperref}
\usepackage[capitalise]{cleveref}
\usepackage{textcomp} 

\usepackage{epstopdf}
\usepackage{cases}
\usepackage{amssymb}
\usepackage{multirow}
\usepackage{siunitx}
\usepackage{cases}
\usepackage{tabularx}
\usepackage{nomencl}
\usepackage{verbatim}
\usepackage[top=1.1in, bottom=1.1in, left=1.0in, right=1.0in]{geometry}
\usepackage[utf8]{inputenc}
\usepackage{tikz}
\usepackage{comment}
\usepackage{lineno,hyperref}
\modulolinenumbers[5]
\usepackage{booktabs}

\begin{document}


\title{Radial thermal rectification in the concentric silicon ring from ballistic to diffusive regime}
\author[add1]{Chuang Zhang}
\ead{zhangcmzt@hust.edu.cn}
\author[add1]{Songze Chen\corref{cor1}}
\ead{jacksongze@hust.edu.cn}
\author[add1]{Zhaoli Guo\corref{cor1}}
\ead{zlguo@hust.edu.cn}
\cortext[cor1]{Corresponding author}
\address[add1]{State Key Laboratory of Coal Combustion, School of Energy and Power Engineering, Huazhong University of Science and Technology,Wuhan, 430074, China}
\date{\today}

\begin{abstract}
The radial thermal rectification in the concentric silicon ring from ballistic to diffusive regime is investigated based on the phonon Boltzmann transport equation.
In the ballistic and diffusive limits, the analytical solutions prove that there is no thermal rectification.
In the ballistic-diffusive regime, the heat flux prefers to flow from the inner boundary to the outer boundary.
Furthermore, as the characteristic length (the distance between two circular boundaries) increases from tens of nanometers to tens of microns, the thermal rectification ratio enhances first and then fades away gradually.
It attributes to that as the direction of the temperature gradient changes, the average phonon mean free path changes.
The difference of the average phonon mean free path finally leads to the change of the heat flux or thermal conductivity.
As the temperature decreases, the maximum thermal rectification ratio decreases.
In addition, as the radius ratio between the inner and outer boundary increases, the thermal rectification ratio decreases for a given characteristic length.

\end{abstract}


\maketitle
\section{Introduction}

Thermal rectification~\cite{starr1936,terraneo2002a,RevModPhysLibaowen} is a kind of anomalous heat transfer phenomenon, in which the heat flux prefers to flow in one direction with higher thermal conductivity.
It has been got much attention since its first experimental observation by Starr~\cite{starr1936}.
A lot of studies, including the numerical simulations or experimental measurements~\cite{RevModPhysLibaowen,yang2012a,liu2019a,roberts2011a}, have been done for better understanding the underlying thermal transport mechanisms.

In the past two decades, most attention has been paid to the phonon management on energy transport
in nanoscale thermal systems~\cite{RevModPhysLibaowen}, such as thermal diodes~\cite{li_thermal_2004} or thermal logic gates~\cite{PhysRevLett.99.177208}.
Theoretical model was firstly developed by Terraneo et al.~\cite{terraneo2002a} for thermal rectifier in $2002$, in which the rectifying effect is obtained by acting on the parameters which control the nonlinearity of the lattice.
Although this model is far away from a realistic implementation, nevertheless, it opens the possibility to propose thermal devices which may have practical relevance.
Li et al.~\cite{li_thermal_2004} demonstrates a thermal diode model that works in a wide range of system parameters by coupling two nonlinear one dimensional lattices.
After above theoretical predictions, in $2006$, Chang et al.~\cite{chang_solid-state_2006} demonstrated nanoscale solid-state thermal rectification, in which high-thermal-conductivity carbon and boron nitride nanotubes were mass-loaded externally and inhomogeneously with heavy molecules.
The experiment resulting nanoscale system yields asymmetric axial thermal conductance with greater heat flow in the direction of decreasing mass density.
Along this line, many simulation investigations and experiments~\cite{wang2017} have focused on various asymmetric nanostructures~\cite{yang2012a,liu2019a,yousefzadi_nobakht_thermal_2018} or lattice system~\cite{ai2011a,ai2011,zhong2009} including mass graded, shape changing (tapered, tailored), such as carbon nanocone~\cite{yang_carbon_2008}, single-walled carbon nanohorns~\cite{wu2008}, graphene nanoribbons~\cite{hu2009,yang_thermal_2009,wang2014}, graphene nanojunctions~\cite{ouyang2010}.
Some normal mechanisms~\cite{RevModPhysLibaowen,liu2019a} are usually used to explain the thermal rectification.
One of them is the phonon spectra of two connecting materials or asymmetric nanostructures.
The calculation of vibrational density of states indicates that the phonon spectra overlap varies by switching the direction of the temperature gradient~\cite{yang_thermal_2009,liu2019a,roberts2011a}.
This difference may be obvious in the nanoscale asymmetry thermal system.
As the phonon mean free path is smaller than the characteristic length of the thermal system, the phonon ballistic scattering~\cite{ouyang2010,miller2009} and local edge scattering~\cite{wang2017} also become important factors for the thermal conductivity.
The asymmetry phonon scattering finally leads to asymmetry heat conduction.

Apart from these mechanisms, another necessary condition~\cite{go2010} for thermal rectification is the thermal conductivity of the material or structure should be a function of both physical space and temperature and nonseparable, which is also one of the mechanisms to explain the thermal rectification in bulk materials.
Peyrard~\cite{peyrard2006} and Dames~\cite{dames2009} note that thermal rectification can be realized in bulk materials by selecting materials with the suitable properties or different temperature dependent thermal conductivity.
Except theoretical analysis, experimental progress in bulk materials is also made by Kobayashi et al.~\cite{kobayashi2009} and Sawaki et al.~\cite{sawaki2011}.
The former prepared an oxide thermal rectifier made of two cobalt oxides with different thermal conductivities~\cite{kobayashi2009}, while the latter investigated thermal rectification in a bulk material with a pyramid shape to elucidate shape dependence of the thermal rectification~\cite{sawaki2011}.

Except the heat transfer in the nanoscale thermal systems or bulk materials, as phonon transports from ballistic to diffusive regime, thermal rectification may happen, too.
The phonon mean free path in silicon or graphene usually ranges $3-4$ orders of magnitudes.
It indicates that even for a given characteristic length, phonon transports crossover from ballistic to diffusive regime~\cite{RevModPhys.90.041002,RevModPhysLibaowen}.
For many asymmetry geometries, as the heat transfer is in the ballistic-diffusive regime, the thermal conductivity usually is dependent of both the spatial space and temperature~\cite{cahill2003nanoscale,cahill2014nanoscale} and nonseparable~\cite{go2010}.
In addition, many previous studies~\cite{RevModPhys.90.041002,cahill2003nanoscale,cahill2014nanoscale} show that the length-dependent thermal conductivity changes rapidly as the characteristic length is comparable to the phonon mean free path.
Hence, it has great potential to realize thermal rectification in the ballistic-diffusive regime.

One of widely used methods to realize the thermal rectification in this regime is changing the spatial configurations of the thermal systems.
Wang et al.~\cite{wang_monte_2010} investigates the phonon transport in single silicon nanowires with variable cross-section.
The results show that the tapered cross-section nanowires can decelerate thermal flux and that the incremental cross-section one has the opposite influence.
However, the simulations are limited below $100$nm.
D. Jou et al.~\cite{criado-sancho2013,criado-sancho2013a} analyses the thermal rectification in inhomogeneous or composite thermal systems from macro-to-nanoscale~\cite{carlomagno2016}, for example the nanoporous/bulk silicon devices~\cite{criado-sancho2012}.
It attributes to that the thermal conductivity decreases significantly as the characteristic length of the thermal system decreases from macro-to-nanoscale and the phonon boundary scattering increases.
Arora et al.~\cite{arora2017} studies the thermal rectification in a selectively restructured graphene by introducing vacancy defects in a potion of graphene.
As a result, they find that the thermal rectification is mainly a function of length of defective and nondefective regions and volume percentage of defect, and it is mostly independent of defect size.
A longer (of the order of $10\mu$m) nondefective side, coupled to a shorter (of the order of $100$nm) defective side, can lead to large thermal rectification.
However, the manufacturing process of the restructured graphene materials is more expensive and complex compared to the silicon.

In this study, the radial thermal rectification in the homogeneous concentric silicon ring from ballistic to diffusive regime is investigated, which is motivated by the nonuniform radial thermal conductivity~\cite{yang_nanoscale_2015,li2019} and radial thermal rectification in graphene~\cite{yousefi2019} and helium II~\cite{saluto2018}.
The paper is organized as follows. The Sec.~\ref{sec:method} we introduce the schematic of the concentric silicon ring and the basic theory of the phonon Boltzmann transport equation (BTE). In Sec.~\ref{sec:results}, the numerical results about the radial thermal rectification are predicted and some analysis are discussed. Finally, a conclusion is made in Sec.~\ref{sec:conclusion}.

\section{Structure and phonon BTE}
\label{sec:method}
\begin{figure}
 \centering
 \includegraphics[scale=0.3,viewport=150 0 750 600,clip=true]{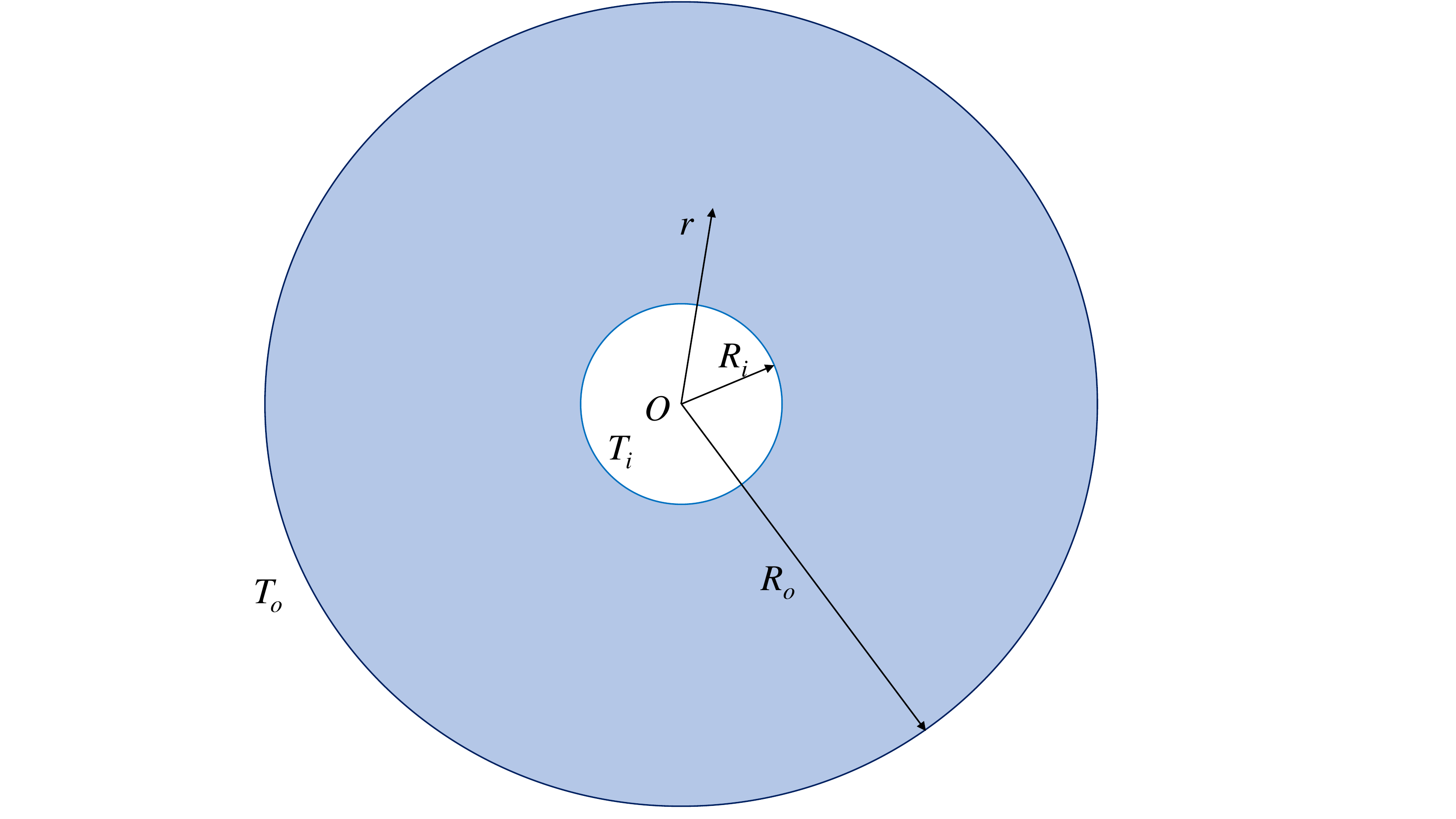}
 \caption{Schematic of the concentric silicon ring. The characteristic length of the thermal system is $L=R_o-R_i$.}
 \label{silicondiskpdf}
\end{figure}

Figure~\ref{silicondiskpdf} shows the simulated thermal system of silicon including two concentric circular boundaries with different radii of $R_i$ and $R_o$, where $R_i<R_o$.
In what follows, the characteristic length of the thermal system is defined as $L=R_o-R_i$.
The temperature of the inner and outer boundaries is fixed at $T_i=T_0 \left(1+ \Delta/2 \right)$ and $T_o=T_0 \left(1 - \Delta/2 \right)$, respectively, where $T_0$ and $\Delta$ are the average temperature and the normalized temperature difference between two boundaries.
As $\Delta \neq 0$, there is temperature gradient along the radial direction.
The total heat flux across the circle with radius $r$ is $Q(r)= 2\pi r \bm{q} \cdot \mathbf{n}$, where $R_i \leq  r \leq R_o $, $\bm{q}$ is the heat flux, $\mathbf{n}$ is the normal unit vector along the radial direction from the inner boundary to the outer boundary.
At steady state, $Q$ is a constant due to the energy conservation.
As $\Delta >0$, the heat flux flows from the inner to the outer boundary and the associated macroscopic variables $W$ ($T,~Q,~\bm{q}$, etc) are labeled as '$W_+$'. As $\Delta<0$, the heat flux flows in the opposite direction and the associated macroscopic variables are labeled as '$W_-$'.
According to previous studies, as $|\Delta|$ increases, thermal rectification~\cite{li_thermal_2004,RevModPhysLibaowen} may happen, i.e., $|Q_+| \neq |Q_-|$.

In order to investigate the thermal rectification in silicon from tens of nanometers to tens of microns, numerical simulations are implemented based on the phonon Boltzmann transport equation (BTE) under the single-mode relaxation time approximation~\cite{ChenG05Oxford,kaviany_2008,MazumderS01MC}, i.e.,
\begin{equation}
\bm{v} \cdot \nabla f  = \frac { f^{0}({T}_{\text{loc}})-f }{\tau(T) },
\label{eq:BTE11}
\end{equation}
where $f=f(\bm{x},\bm{s},\omega,p)$ is the phonon distribution function with the space vector $\bm{x}$, unit directional vector $\bm{s}$ in 3D coordination, phonon angular frequency $\omega$ and polarization $p$.
$\bm v=\nabla_{\bm{K}} {\omega}$ is the group velocity calculated by the phonon dispersion, where $\bm{K}$ is the wave vector and assumed to be isotropic.
The optical phonon branches are not considered due to the small contribution to the thermal conduction.
The approximate quadratic polynomial dispersions~\cite{pop2004analytic} are used to represent the dispersion relation~\cite{brockhouse1959lattice} of the acoustic phonon branches in monocrystalline silicon.
$\tau=\tau(T)$ is the effective relaxation time, a combination of various phonon-phonon intrinsic scattering (include impurity, N and U scattering) mechanisms based on the Matthiessen's rule~\cite{holland1963analysis,kaviany_2008}, where $T$ is the temperature and will be discussed later.
Here, the experimental formulas of the impurity, N and U scattering are used, which can refer to Ref~\cite{terris2009modeling}.
$f^{0}$ is the local equilibrium state with pseudo-temperature ${T}_{\text{loc}}$, satisfying the Bose-Einstein distribution~\cite{ChenG05Oxford,kaviany_2008}, i.e.,
$$f^{0} (T)= f_{\text{BE}} (T)= \frac{1}{ \exp(\hbar\omega/k_{B}T)-1  } ,$$
where $\hbar$ is the Planck's constant divided by $2\pi$, $k_B$ is the Boltzmann constant.
The pseudo-temperature ${T}_{\text{loc}}$ is introduced to ensure the energy conservation of the scattering term, i.e.,
\begin{equation}
\sum_{p}\int \int_{4\pi}  \frac{\hbar \omega D}{4\pi}   \frac { f^{0}(T_{\text{loc}})-f }{\tau(T)} d{\Omega}d{\omega}=0,
\label{eq:scatteringterm}
\end{equation}
where $D$ is the phonon density of state, $d{\Omega}$ and $d{\omega}$ are the integral over the solid angle space and the frequency space, respectively.
In systems out of equilibrium, the temperature $T$ can be defined in terms of an equilibrium distribution with the same energy density, namely, equivalent equilibrium temperature~\cite{ChenG05Oxford}, i.e.,
\begin{equation}
\sum_{p}\int \int_{4\pi} \frac{\hbar \omega D}{4\pi} \left( f_{\text{BE}}(T)-f  \right)  d{\Omega}d{\omega}=0.
\label{eq:thermodynamicterm}
\end{equation}
The heat flux $\bm{q}$ is calculated by
\begin{equation}
\bm{q} =\sum_{p}\int \int_{4\pi} \bm{v}  \hbar \omega f D/4{\pi} d{\Omega}d{\omega}.
\label{eq:heatflux}
\end{equation}

An implicit synthetic scheme~\cite{ADAMS02fastiterative} is used to solve Eq.~\eqref{eq:BTE11}, which can refer to Ref~\cite{ZHANG20191366}.
To ensure the numerical accuracy, $30164 \times  1152 \times  40$ cells are used to discrete the physical space, solid angle space and frequency space, respectively.
In addition, the isothermal thermalizing boundary conditions~\cite{li2019} are implemented on the inner and outer boundaries.

\section{Results and discussions}
\label{sec:results}

Previous work~\cite{li2019} based on phonon gray model shows that the radial heat transport is dominated by two parameters including the radius ratio of the two concentric boundaries (${R_i}/{R_o}$) and the ratio of the phonon mean free path to the characteristic length.
Hence, the thermal rectification in the ballistic-diffusive regime with different characteristic length $L$, temperature range ($T_0,\Delta$) and $R_i/R_o$ are investigated.

First, the thermal rectification in the ballistic and diffusive limits are derived theoretically.
The radial local thermal conductivity $k$ is introduced and defined as
\begin{align}
k   &= \frac{-Q}{2\pi r \frac{dT}{dr} }, \label{eq:conductivity}  \\
\frac{dT}{Q}  &= -\frac{1}{2\pi k} d(\ln{r})  .  \label{eq:integral}
\end{align}
where $k=k(R_i/R_o,L,r,T,T_0, \Delta)$ is dependent of the geometry $(R_i/R_o,~L)$ and temperature range $(T_0,~\Delta)$ of the thermal system as well as the spatial position $r$ and temperature $T(r)$.
As the characteristic length is much larger than the phonon mean free path and the heat transfer is in the diffusive regime, the radial thermal conductivity is only dependent of the local temperature for a given system geometry, i.e., $k=k(T)$~\cite{rudramoorthy2010heat}.
Then taking an integral of Eq.~\eqref{eq:integral} from $r=R_i$ to $r=R_o$ as $\Delta>0$ or $\Delta<0$, respectively, we can derive~\cite{go2010} (as $\Delta>0$, subscript of all macroscopic physical quantities plus one '+' symbol, else '-')
\begin{align}
\frac{1}{Q_+}  \int_{T_i}^{T_o}  k(T)  dT &=  \int_{R_i}^{R_o}  -\frac{1}{2\pi } d(\ln{r}), ~~\Delta>0,  \\
\frac{1}{Q_-}  \int_{T_i}^{T_o}  k(T)  dT &=  \int_{R_i}^{R_o}  -\frac{1}{2\pi } d(\ln{r}), ~~\Delta<0, \\
\Longrightarrow   |Q_+| &= |Q_-|.
\label{eq:integraldiffusive}
\end{align}
As the characteristic length is much smaller than the phonon mean free path and the heat transfer is in the ballistic regime, in which there is rare phonon-phonon intrinsic scattering.
For arbitrary phonon coming from one boundary, it incidents into other boundaries directly without energy and momentum change.
The total heat flux can be calculated by~\cite{li2019,olfe1968}
\begin{align}
Q_{+} &= F(T_i)-F(T_o) ,~\Delta>0,  \\
Q_{-} &= F(T_i)-F(T_o) ,~\Delta<0,  \\
\Longrightarrow   |Q_+| &= |Q_-|,
\label{eq:integralballistic}
\end{align}
where
\begin{align}
F(T)=0.5 \pi R_i   \sum_{p}\int \hbar \omega D|\bm{v} |  f_{\text{BE}}(T)   d{\omega} .
\end{align}
In addition, it can be observed that the total heat flux in the ballistic limit is nothing to do with the phonon mean free path.
In a word, there is no thermal rectification in the diffusive and ballistic limits.

\begin{figure}
     \centering
     \subfloat[]{\label{rectification2}\includegraphics[width=0.4\textwidth]{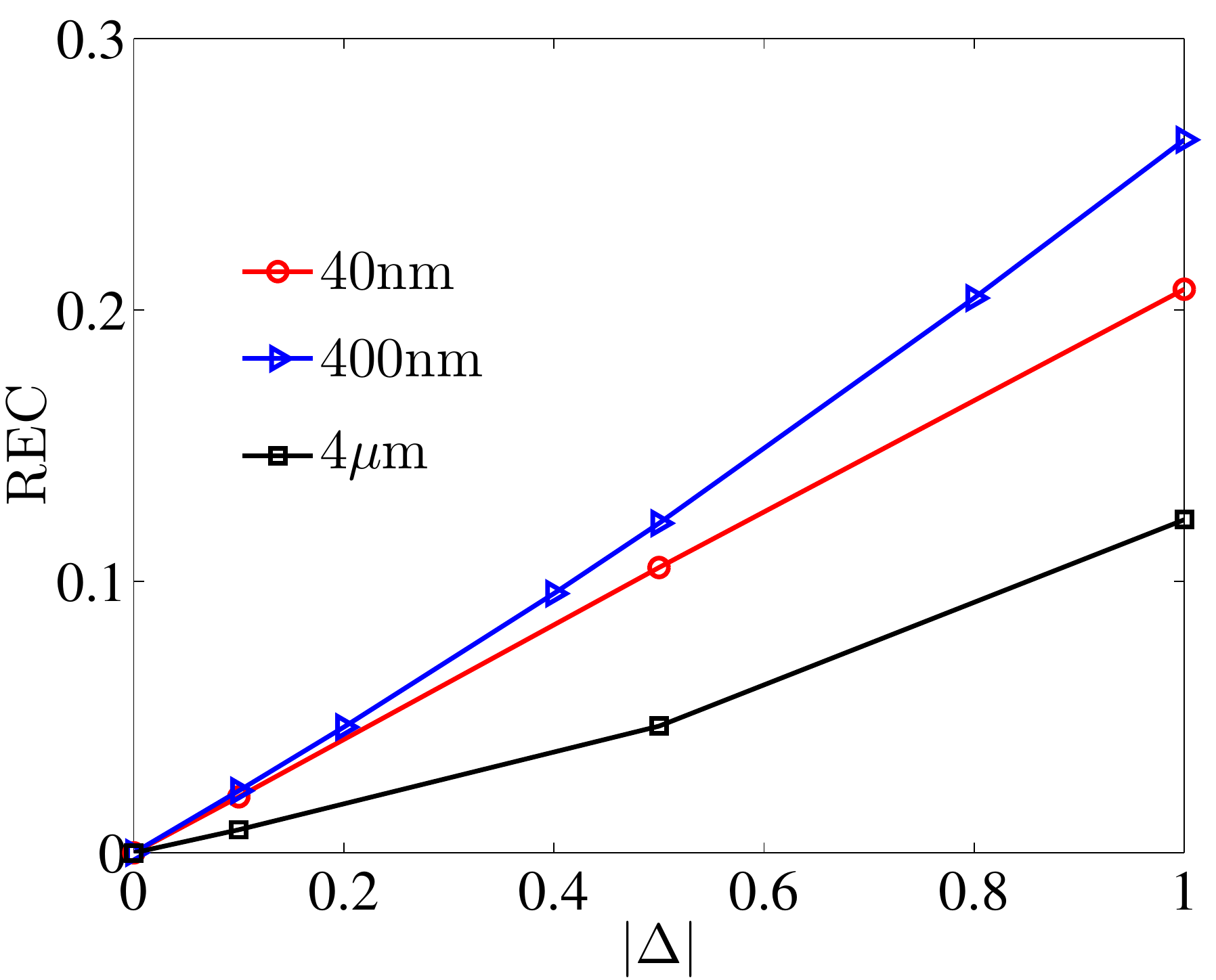}}~~
     \subfloat[]{\label{systemsizeDT}\includegraphics[width=0.4\textwidth]{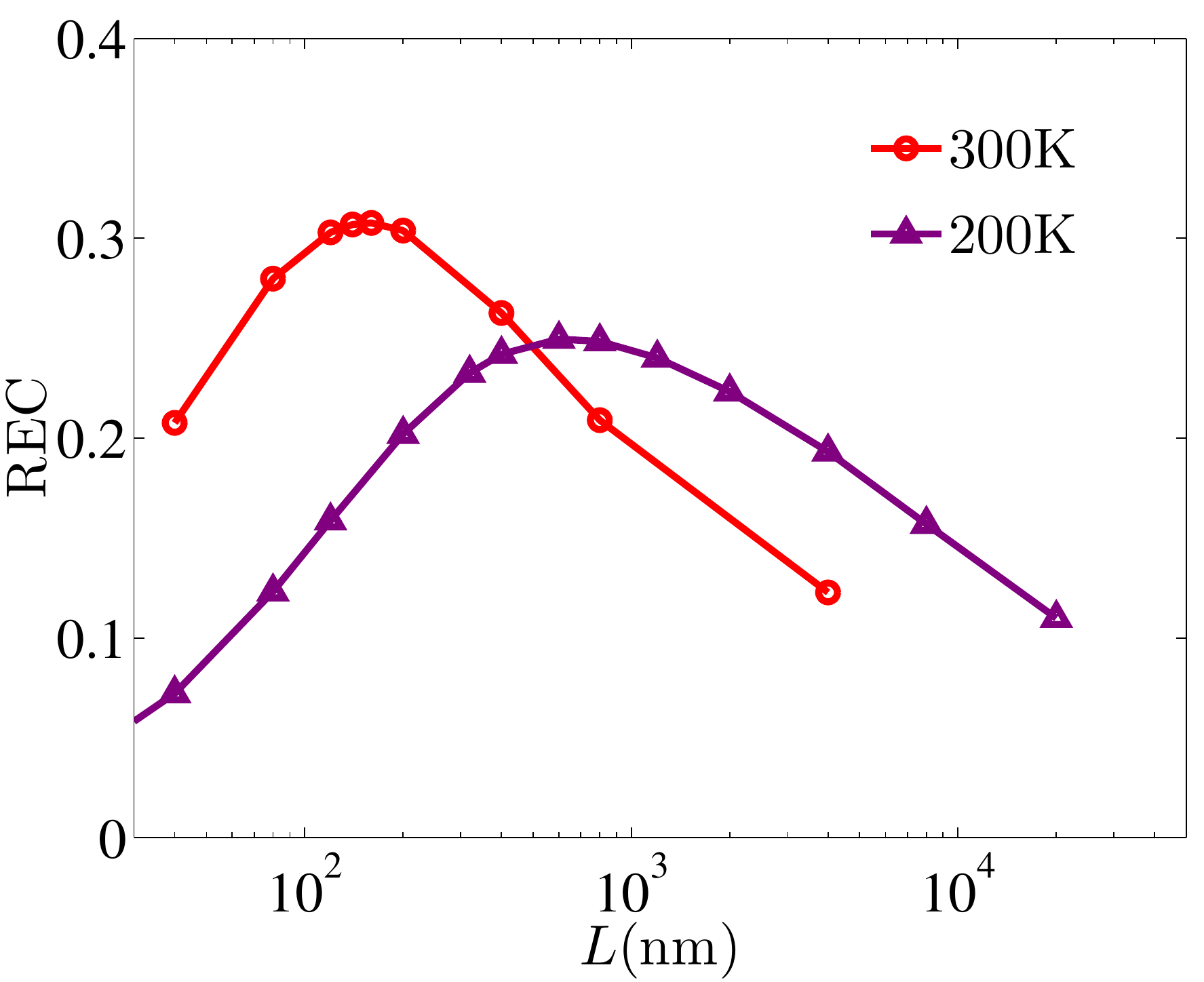}}~~\\
     \caption{(a) The distributions of the thermal rectification ratio $\text{REC}$ at different $|\Delta|$ with different characteristic length $L=R_o-R_i$, where $T_0=300\text{K}$, $R_i/R_o=0.2$. (b) The distributions of the thermal rectification ratio with different temperature $T_0$ ($300\text{K}$, $200\text{K}$) and characteristic length $L=R_o-R_i$, where $|\Delta|=1.0$, $R_i/R_o=0.2$.}
     \label{systemsizeDT22}
\end{figure}

In the ballistic-diffusive regime, the thermal rectification is more complicated and there is no analytical solutions.
Figure~\ref{rectification2} shows the thermal rectification ratio ($\text{REC}$) at different $|\Delta|$ ($0.001-1.0$) with different characteristic length $L$, where $R_i/R_o=0.2$ and the thermal rectification ratio is defined as
\begin{align}
\text{REC}= \frac{ \left|Q_+ \right| -\left| Q_- \right|}{ \left|Q_-\right|}.
\end{align}
It can be observed that as $|\Delta|$ increases from $0.001$ to $1.0$, the thermal rectification ratio $\text{REC}$ increases whatever the characteristic length.
In addition, the heat flux prefers to flow from the inner to the outer boundary, which is opposite to the previous results in graphene predicted by the molecular dynamics~\cite{yousefi2019}.
Unlike previous molecular dynamics simulations in which the phonon density of states along the radial direction are nonuniform at the nanoscale, in the BTE scale, the asymmetric atomic details and phonon wave nature are not accounted and the phonon density of states are constant for a given phonon frequency and polarization.

Furthermore, it is interesting to find that for a given temperature difference $|\Delta|$, the thermal rectification ratio with $L=400$nm is larger than that as $L=40$nm or $L=4\mu$m, as shown in~\cref{rectification2}.
According to that there is little thermal rectification in the ballistic and diffusive limits, we can concluded that for a given $|\Delta|$, as the characteristic length increases from $40$nm to $4\mu$m, there is at least an extremum value of the thermal rectification ratio.
In order to predict the extremum value more accurately, numerical simulations are implemented with different characteristic length $L$ from tens of nanometers to tens of microns, where $|\Delta|=1.0$, $R_i/R_o=0.2$, as shown in~\cref{systemsizeDT} and Table.~\ref{rectificationdata}.
It can be observed that as $T_0=300\text{K}$, with the increase of the characteristic length, the thermal rectification ratio increases gradually till maximum value, then decreases.
The maximum value reaches as the characteristic length is $140$nm$-160$nm.
\begin{table}
\caption{The specific numerical data of the thermal rectification ratio ($\text{REC}$) with different $T_0$, $L$ and $R_i/R_o$, where $|\Delta |=1.0 $, as shown in~\cref{systemsizeDT} and~\cref{GDT3}.}\vskip 0.2cm
\centering
\begin{tabular}{|*{6}{c|}}
\hline
\multicolumn{4}{|c|}{ $T_0=300 \text{K}$ }  & \multicolumn{2}{c|}{ $T_0=200 \text{K}$ } \\
\hline
\multicolumn{2}{|c|}{ $R_i/R_o=0.2$ }  & \multicolumn{2}{c|}{ $R_i/R_o=0.6$ }  & \multicolumn{2}{c|}{ $R_i/R_o=0.2$ } \\
\hline
 $L$(nm) & $\text{REC}$ & $L$(nm) & $\text{REC}$ & $L$(nm) & $\text{REC}$  \\
\hline
 40  & 0.208  & 20  & 0.123  & 80 &    0.123  \\
  \hline
120  &   0.303 & 40 & 0.155&   120 & 0.159  \\
 \hline
140  &   0.307 & 60 & 0.162 &  200  & 0.202  \\
 \hline
160  &   0.308 & 80 & 0.159 &  400 & 0.242 \\
 \hline
200  &   0.304  & 100 & 0.155 & 600 & 0.250  \\
 \hline
400 &   0.263  & 200 & 0.127 & 800 & 0.248  \\
 \hline
 800  &   0.209 & 400 & 0.09  &1200 & 0.240  \\
 \hline
 4000  &   0.123 & 2000 & 0.04   & 2000 & 0.223 \\
 \hline
\end{tabular}
\label{rectificationdata}
\end{table}
\begin{figure}
     \centering
     \subfloat[]{\label{idea1}\includegraphics[width=0.6\textwidth]{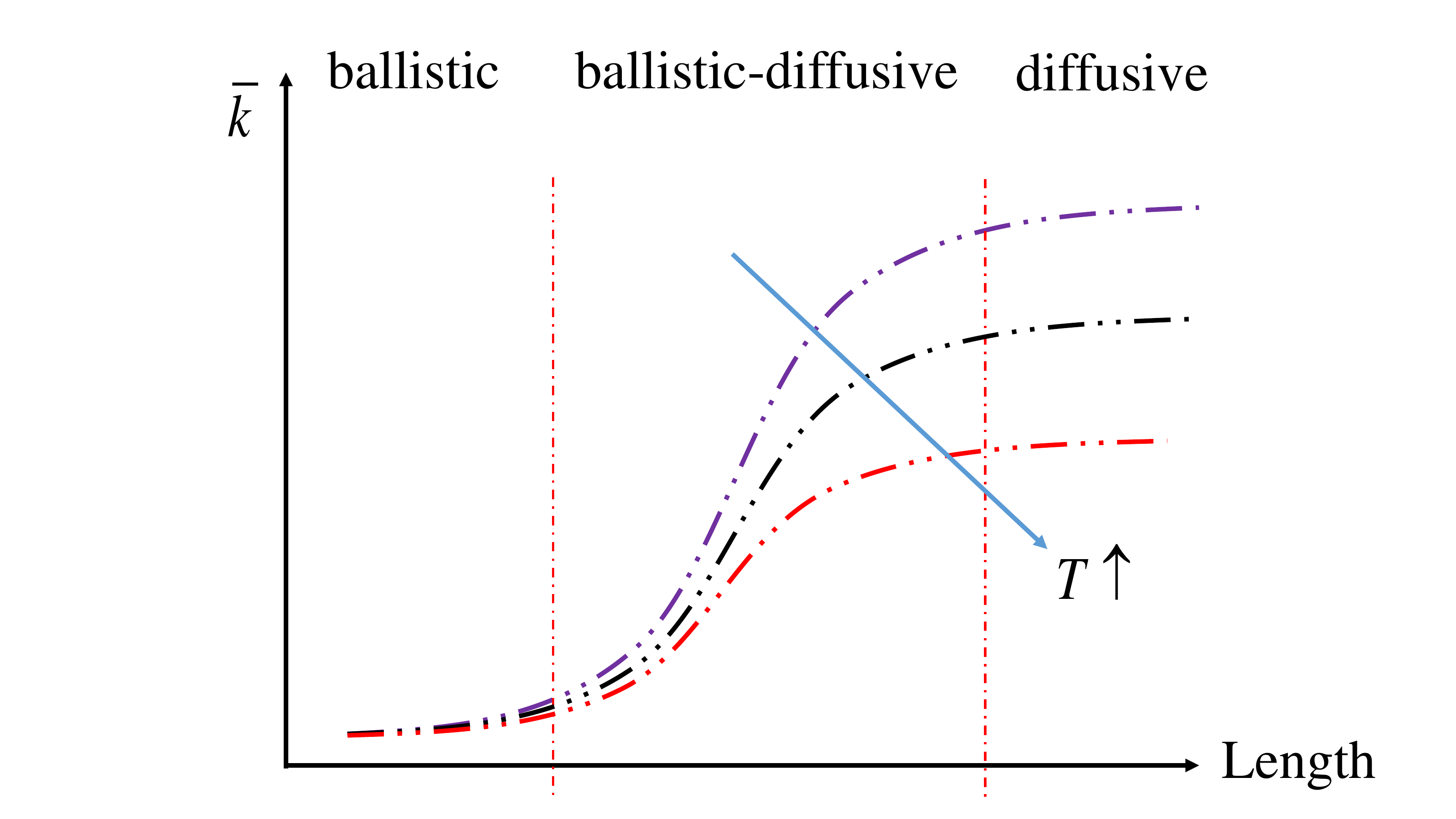}}~~\\
     \subfloat[]{\label{idea2}\includegraphics[width=0.6\textwidth]{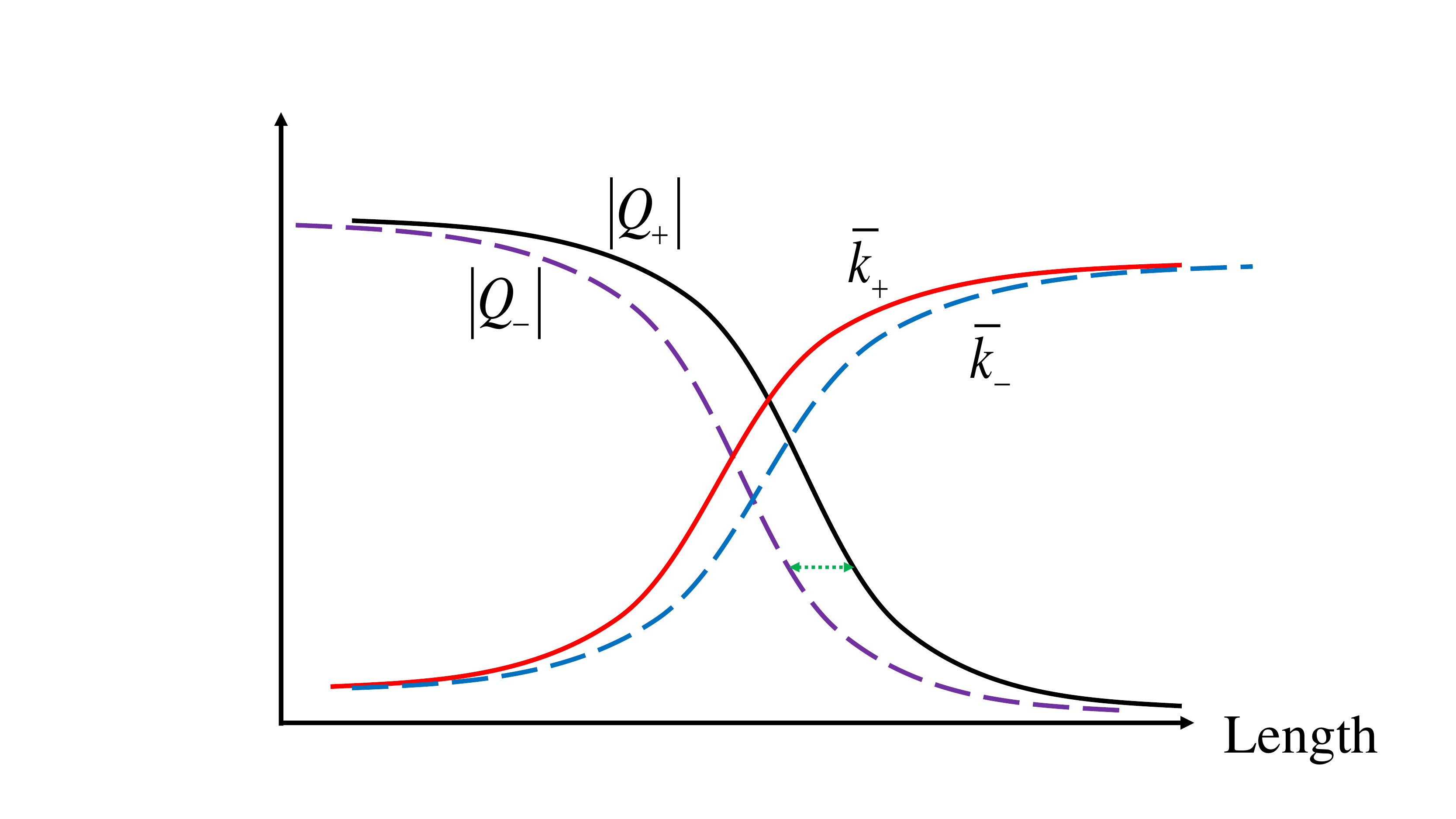}}~~\\
     \caption{(a) Schematic of the distributions of the effective thermal conductivity $\overline{k}$ with different characteristic length $L$ for a given temperature $T$ in silicon~\cite{li2019,ChenG05Oxford,terris2009modeling}. (b) A schematic of how to realize the thermal rectification in the ballistic-diffusive regime. If we can change the phonon mean free path as the direction of the temperature gradient changes, the profiles of the length-dependent heat flux $|Q|$ or effective thermal conductivity $\overline{k}$ may be stretched or shrunk along the horizontal direction. $+$ and $-$ represent the heat flows from the inner to the outer or in the opposite direction, respectively. }
     \label{rectification}
\end{figure}

We suppose that one of the reasons for these phenomena is the size effects, namely, the effective thermal conductivity $\overline{k}$ in silicon decreases as the system size decreases~\cite{cahill2014nanoscale,cahill2003nanoscale}.
Besides, the effective thermal conductivity profiles change most rapidly as the system size is comparable to the phonon mean free path~\cite{li2019,ChenG05Oxford}, as shown in~\cref{idea1}.
On the other hand, for a given system size, as the temperature increases, the phonon-phonon intrinsic momentum destroying scattering happens frequently so that the effective thermal conductivity decreases.
Hence, in the ballistic-diffusive regime, both the phonon boundary scattering and the phonon-phonon intrinsic scattering play an important role on the heat transfer.
Furthermore, for the concentric ring geometry, previous studies~\cite{yang_nanoscale_2015,li2019} have proven that even for a given system size and temperature ($T_0,~\Delta \rightarrow 0$), the local radial thermal conductivity is dependent of the radius in the ballistic-diffusive regime, i.e., $$k=k(r).$$
It is totally different from that in the cross-plane heat transfer, in which the local thermal conductivity is independent of the spatial space~\cite{MajumdarA93Film}.
Then as $|\Delta|$ is large, the local radial thermal conductivity is possible to be the function of both the radius $r$ and the temperature $T(r)$ and nonseparable, i.e., $$k=k(T,r)=k(T(r),r).$$
If so, it is possible to realize thermal rectification based on the theory in Ref~\cite{go2010}.
So how to realize the thermal rectification?
As shown in~\cref{idea2}, if we can change the phonon mean free path of the thermal system in the ballistic-diffusive regime, the length-dependent total heat flux $Q$ or effective thermal conductivity $\overline{k}$ profiles may be stretched or shrunk along the horizontal direction.
For a given characteristic length, the small change affects a lot to the heat flux, especially in the ballistic-diffusive regime, which results in $|Q_{+}| \neq |Q_{-}|$. (as $\Delta>0$, subscript of all macroscopic physical quantities plus one '+' symbol, else '-')
Hence, one of the key point to realize the thermal rectification is to make the average phonon mean free path of the concentric silicon ring different as the direction of the temperature gradient changes, i.e., $\overline{\lambda}_{+} \neq \overline{\lambda}_{-}$, which is also the main starting point of this work.

\begin{figure}
     \centering
     \subfloat[]{\label{Temperatureradial}\includegraphics[width=0.46\textwidth]{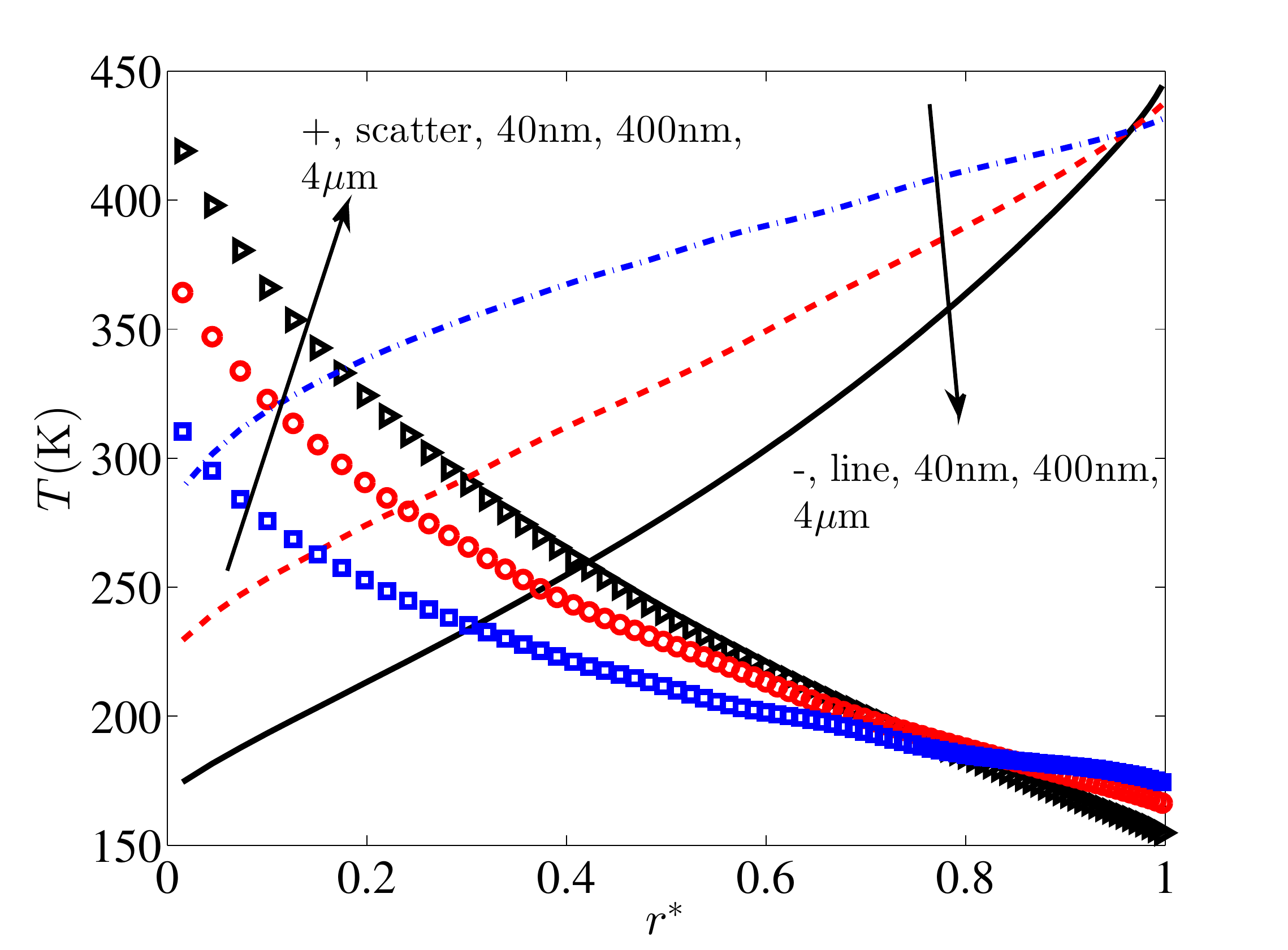}}~~
     \subfloat[]{\label{GDTTL}\includegraphics[width=0.41\textwidth]{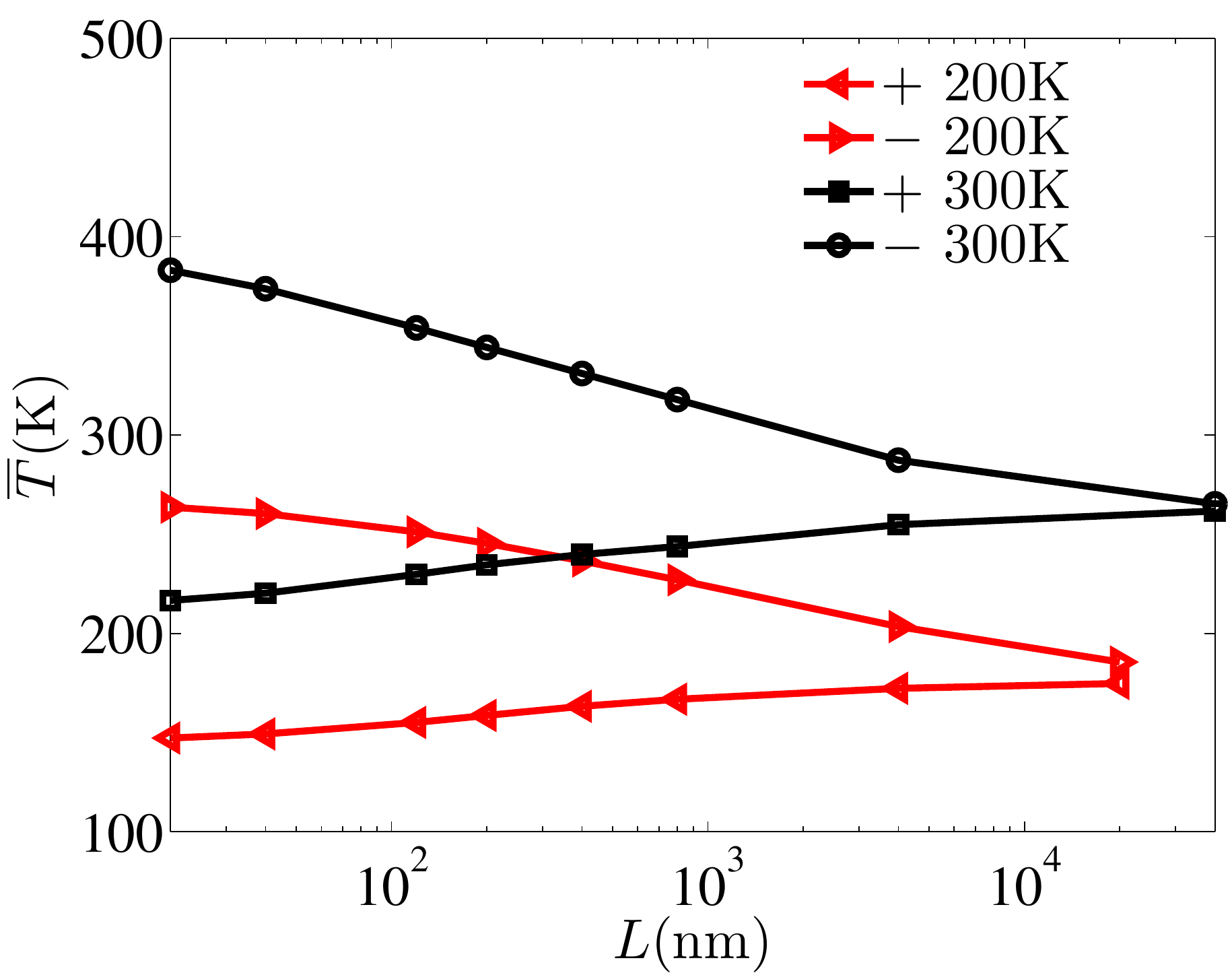}}~~\\
     \caption{$+$ and $-$ represent the heat flows from the inner to the outer or in the opposite direction, respectively. (a) The radial distributions of the temperature with different characteristic length, where $|\Delta|=1.0$, $T_0=300$K, $r^*=\left( \ln(r)-\ln(R_i) \right)/ \left( \ln(R_o)-\ln(R_i) \right)$. (b) The distributions of average temperature $\overline{T}$ with different characteristic length $L=R_o-R_i$ and temperature $T_0$ ($300\text{K}$, $200\text{K}$), where $|\Delta|=1.0$, $ \overline{T}= \frac{ \int_0^1 {T(r^*)} d(r^*)  }{ \int_0^1 d(r^*)  }  $, $R_i/R_o=0.2$. }
     \label{idea}
\end{figure}
In order to prove our guess, the temperature distributions along the radial direction with different characteristic length are predicted, as shown in~\cref{Temperatureradial}, where $T_0=300$K, $|\Delta|=1.0$, $r^*=\left( \ln(r/R_i) ) \right)/ \left( \ln(R_o/R_i) \right)$.
It can be observed that with the increase of the characteristic length, the average temperature $\overline{T}$ increases as $\Delta>0$ but decreases as $\Delta <0$, where $$ \overline{T}= \frac{ \int_0^1 {T(r^*)} d(r^*)  }{ \int_0^1 d(r^*)  }. $$
Previous studies have demonstrated that the average phonon mean free path $\overline{\lambda}(\overline{T})$ in silicon decreases as the temperature increases~\cite{Glassbrenner64conductivity,terris2009modeling,zhang_discrete_2019}, where
\begin{align}
\overline{\lambda}(\overline{T})&=\frac{   \sum_{p} \int { \hbar \omega D \frac{\partial f_{\text{BE}}  }{\partial{T}} }  | \bm{v} | \tau d\omega }  {  \sum_{p} \int { \hbar \omega D \frac{\partial f_{\text{BE}}  }{\partial{T}} } d\omega  }. \label{eq:lambdaeq}
\end{align}
In other words, the average phonon mean free path with $\Delta<0$ is smaller than that with $\Delta >0$.
Usually, for a given characteristic length, the larger the average phonon mean free path in silicon is, the larger the thermal conductivity is, which proves that the heat flux prefers to flow from the inner to the outer side.

\begin{figure}
     \centering
     \includegraphics[width=0.4\textwidth]{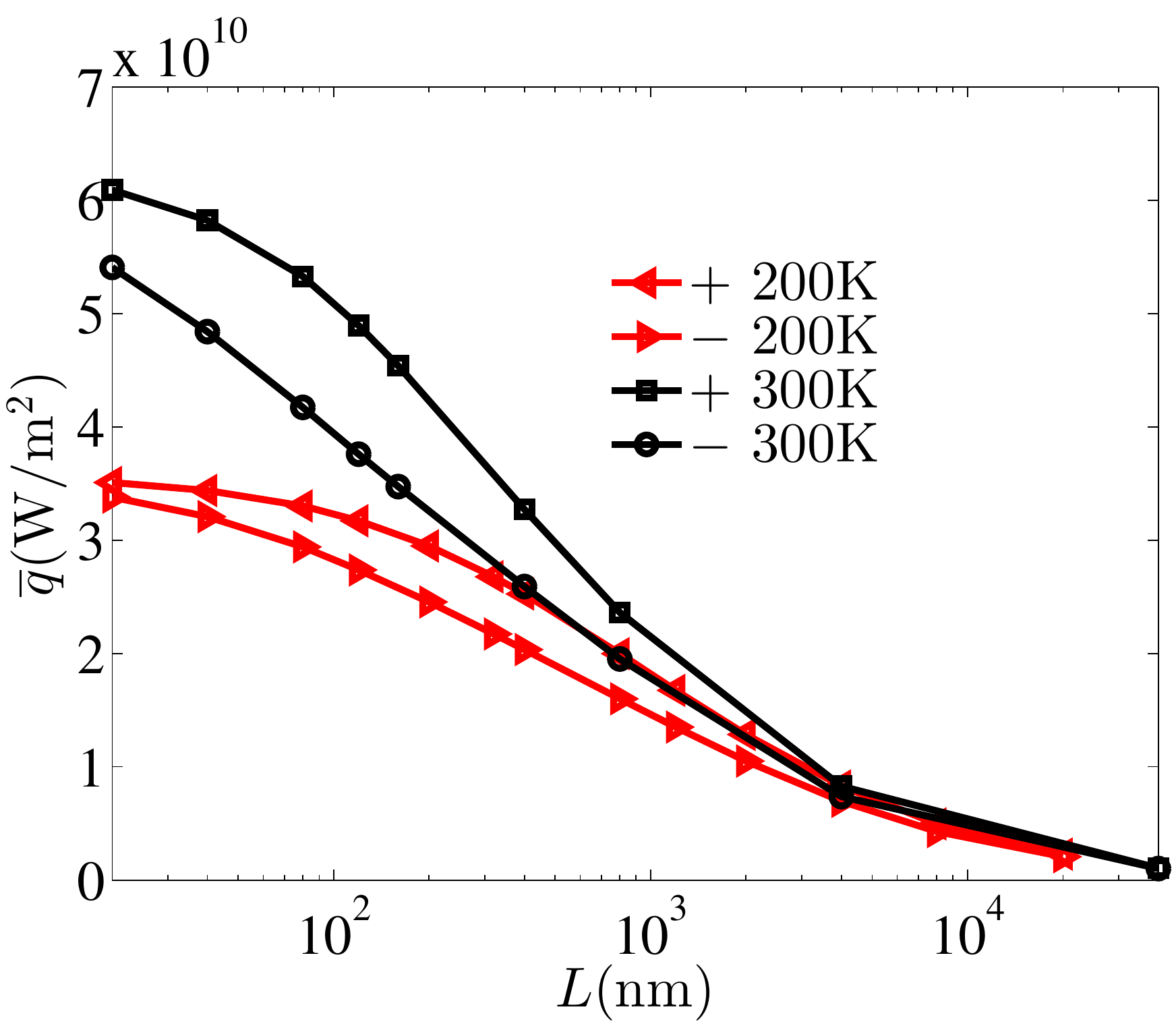}~~\\
     \caption{The distributions of heat flux $\overline{q}$ along the radial direction with different characteristic length $L=R_o-R_i$ and temperature $T_0$ ($300\text{K}$, $200\text{K}$), where $|\Delta|=1.0$, $\overline{q}= \left| \frac{ \int_{R_i}^{R_o}  Q(r)dr  }{ \int_{R_i}^{R_o} 2\pi r  dr } \right|$, $R_i/R_o=0.2$. $+$ and $-$ represent the heat flows from the inner to the outer or in the opposite direction, respectively.}
     \label{GDTQQ}
\end{figure}

The average temperature and phonon mean free path as $\Delta>0$ or $\Delta <0$ are also calculated based on Eq.~\eqref{eq:lambdaeq}.
As shown in~\cref{GDTTL} or Table.~\ref{TaverageTable}, it can be observed that with the increase of the characteristic length, the average temperature increases as $\Delta >0$ while decreases as $\Delta <0$.
As the characteristic length is much large ($L=40\mu$m), the average temperature as $\Delta<0$ is close to that as $\Delta>0$, which goes to the heat transfer in the diffusive regime.
As $T_0=300\text{K}$, $40\text{nm} \leq L \leq 40 \mu$m, we calculate that if $\Delta> 0$, $ 220 \text{K} \leq  \overline{T}_{+} \leq  265 \text{K}$ and $ 232\text{nm} \leq \overline{\lambda}_{+} \leq 300 $nm, else if $\Delta< 0$, $265 \text{K} \leq \overline{T}_{-} \leq 375 \text{K}$ and $152 \text{nm} \leq  \overline{\lambda}_{-} \leq 232 \text{nm}$.
As $L$ increases, $\overline{T}_{-}- \overline{T}_{+}$ decreases gradually and goes to zero, which is consistent with the heat transfer in the diffusive regime.
Figure~\ref{GDTQQ} shows the distributions of the average heat flux $\overline{q}$ with different characteristic lengths, where $\overline{q}= \left| \frac{ \int_{R_i}^{R_o} Q(r)dr  }{ \int_{R_i}^{R_o} 2\pi r  dr } \right|$, $R_i/R_o=0.2$.
It can be observed that as $T_0=300\text{K},~ \Delta>0$, with the increase of the characteristic length crossover from the ballistic to diffusive regime, the heat flux decreases.
And the main drop happens in the ballistic-diffusive regime.
Similar phenomena can be observed as $T_0=300\text{K},~ \Delta<0$.
What's different is that as $\Delta <0$, the average temperature is higher so that $\overline{\lambda}_{-}<\overline{\lambda}_{+}$.
It can be seen that due to $\overline{\lambda}_{-}<\overline{\lambda}_{+}$, with the increase of the characteristic length $L$, the heat flux $\overline{q}_{-}$ decreases dramatically first before $\overline{q}_{+}$.
Besides, $\overline{q}_{-}$ converges first before $\overline{q}_{+}$ as the characteristic length is large enough.
In other words, the numerical profiles are stretched along the horizontal direction so that there is thermal rectification in the ballistic-diffusive regime, which is consistent with our original guess shown in~\cref{idea2}.

\begin{table}
\caption{The specific numerical data of the average temperature along the radial direction ($\overline{T}$) with different $T_0$, $L$ and $R_i/R_o$, where $|\Delta |=1.0 $, $ \overline{T}= \frac{ \int_0^1 {T(r^*)} d(r^*)  }{ \int_0^1 d(r^*)  }  $, as shown in~\cref{GDTTL} and~\cref{RTTL}.}\vskip 0.2cm
\centering
\begin{tabular}{|*{9}{c|}}
\hline
\multicolumn{6}{|c|}{ $T_0=300 \text{K}$ }  & \multicolumn{3}{c|}{ $T_0=200 \text{K}$ } \\
\hline
\multicolumn{3}{|c|}{ $R_i/R_o=0.2$ }  & \multicolumn{3}{c|}{ $R_i/R_o=0.6$ }  & \multicolumn{3}{c|}{ $R_i/R_o=0.2$ } \\
\hline
 $L$(nm) & $\overline{T}_{-}$(K) & $\overline{T}_{+}$(K) & $L$(nm) & $\overline{T}_{-}$(K) & $\overline{T}_{+}$(K) & $L$(nm) & $\overline{T}_{-}$(K) & $\overline{T}_{+}$(K) \\
\hline
 40  & 374  & 220  & 60   & 331&   265 & 40  & 260 & 149  \\
  \hline
120  &   354 & 230  & 100 &  323 & 267  & 120 & 251 &    155 \\
 \hline
200  &   344 & 235 & 200 &  311  & 267  & 200  & 245 &   159 \\
 \hline
 400  &   331 & 240 &400 &  300  & 266 & 400  & 236 &  163 \\
 \hline
 800  &   318 & 244 & 1000 &  287  & 265  & 800  & 227 &  167 \\
 \hline
 4000  &  287 & 255 & 2000 &  280  & 265 & 4000  & 203 &  172 \\
 \hline
 40000  &  265 & 261 & \multicolumn{3}{c|}{  }  & 20000  & 185 &  174 \\
 \hline
\end{tabular}
\label{TaverageTable}
\end{table}

The thermal rectification with different temperature $T_0$ is also investigated, where $R_i/R_o=0.2$ and $|\Delta| =1.0$.
As the temperature decreases from $T_0=300\text{K}$ to $T_0=200\text{K}$, as shown in~\cref{systemsizeDT}, it can be observed that the length dependent thermal rectification ratio profiles move along the positive axis, namely, the increasing of the characteristic length.
There is still a maximum value of the thermal rectification ratio but its associated characteristic length increases.
Besides, as temperature decreases, the maximum thermal rectification ratio decreases.
Similar to the analysis mentioned before, according to Table.~\ref{TaverageTable} and Eq.~\eqref{eq:lambdaeq}, as $T_0=200\text{K}$, $40\text{nm} \leq L \leq 20 \mu$m, if $\Delta> 0$, we predict that $ 149 \text{K} \leq  \overline{T}_{+}  \leq  175 \text{K}$ and $ 420 \text{nm} \leq \overline{\lambda}_{+} \leq 545 $nm, else if $\Delta< 0$, $185  \text{K} \leq \overline{T}_{-} \leq 260 \text{K}$ and $235 \text{nm} \leq  \overline{\lambda}_{-} \leq 385 \text{nm}$.
As $\overline{\lambda}_{-}/L \approx 1$, $\overline{\lambda}_{+}/L \approx 1$, the heat transfer is in the ballistic-diffusive regime.
The different phonon mean free path finally leads to different total heat flux, as shown in~\cref{GDTQQ}.
\begin{figure}
     \centering
     \subfloat[]{\label{GDT3}\includegraphics[width=0.4\textwidth]{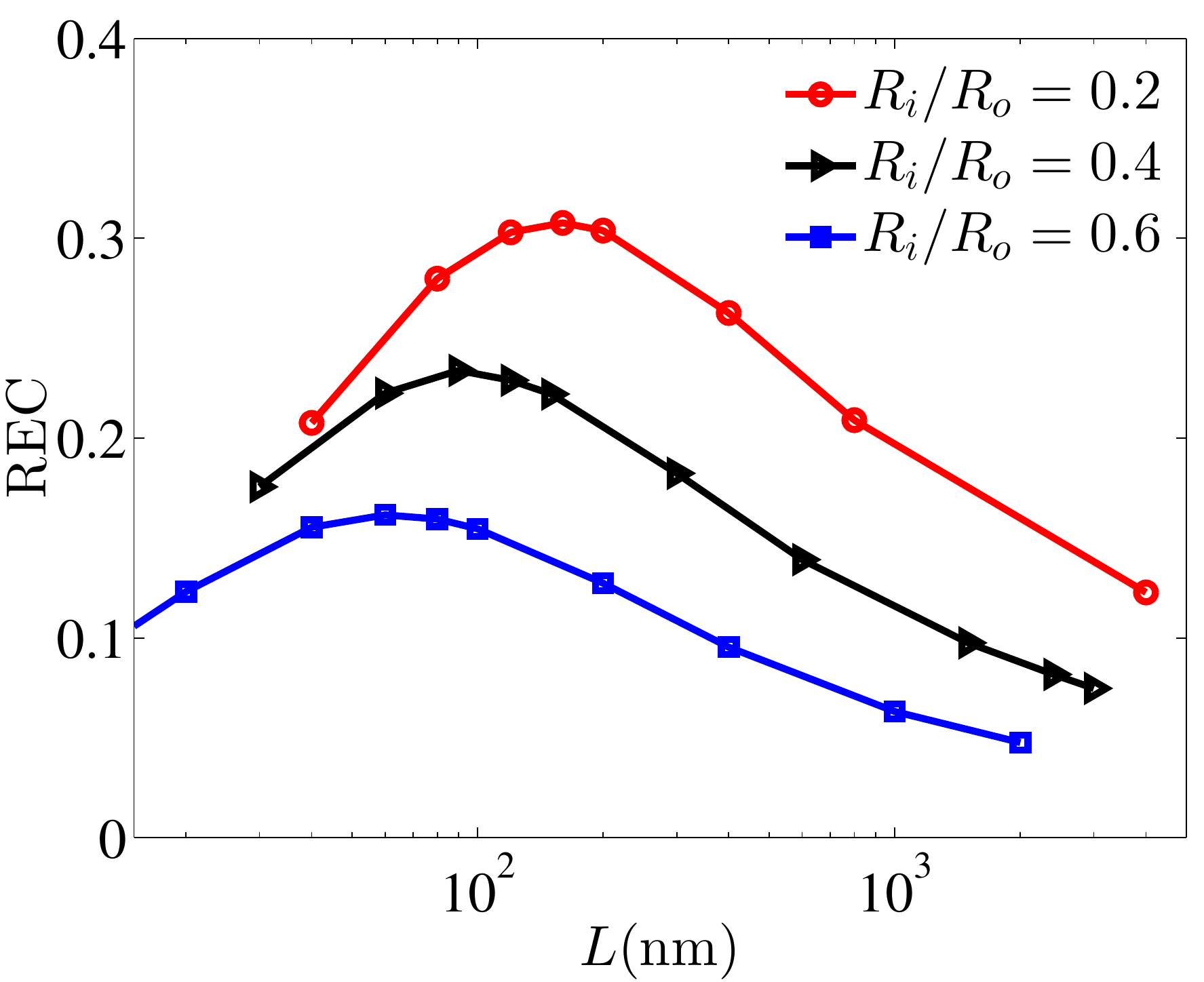}}~~
     \subfloat[]{\label{RTTL}\includegraphics[width=0.4\textwidth]{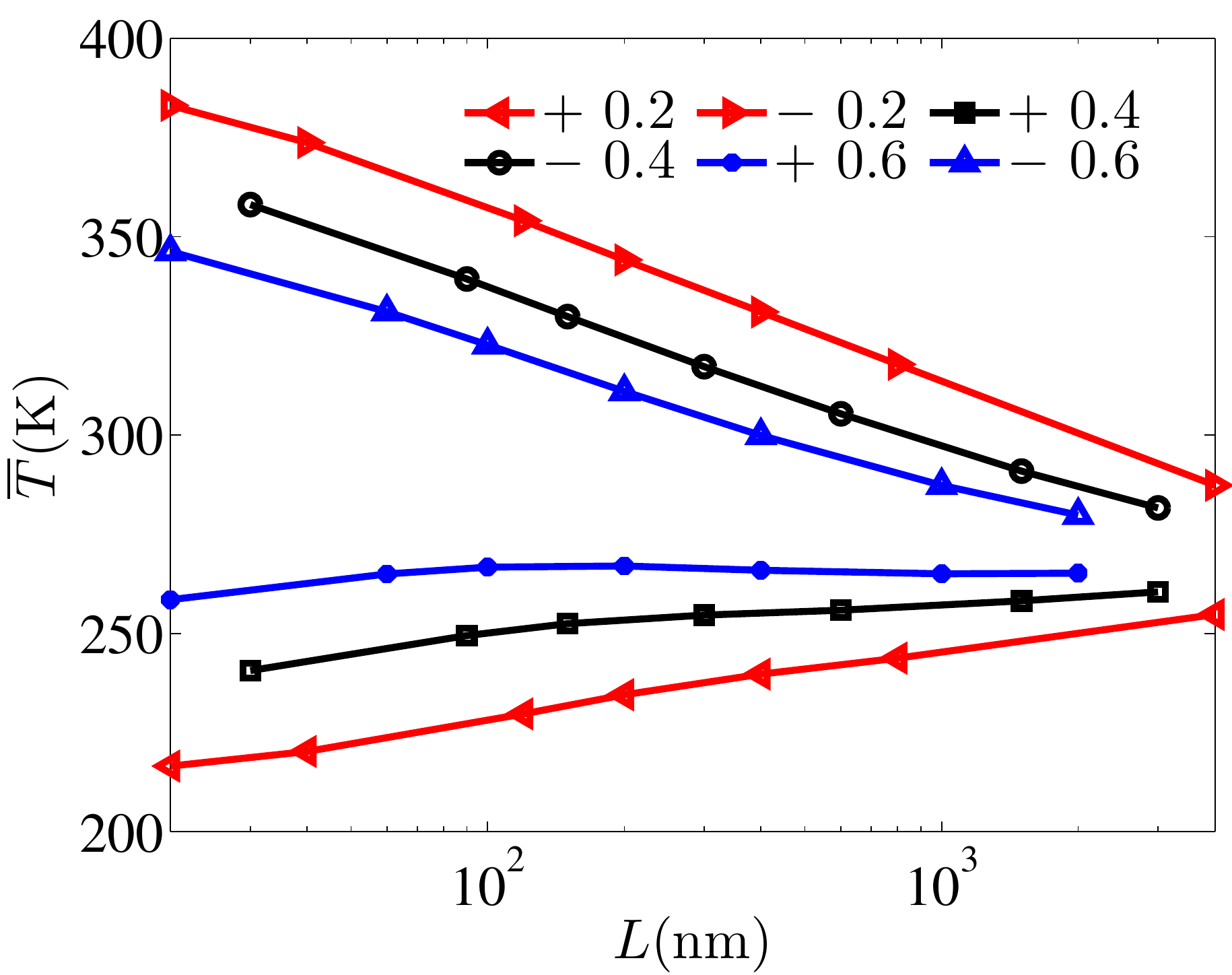}}~~\\
     \caption{The distributions of the thermal rectification ratio $\text{REC}$ (a) and the average temperature in the domain $\overline{T}$ (b) with different characteristic length $L=R_o-R_i$ and $R_i/R_o$ ($0.2,~0.4,~0.6$), where $T_0=300\text{K},~|\Delta|=1.0$, $\overline{T}= \frac{ \int T(r^*)d(r^*)  }{ \int d(r^*) } $. $+$ and $-$ represent the heat flows from the inner to the outer or in the opposite direction, respectively. }
     \label{RIOGDT}
\end{figure}

Except the temperature range ($T_0,~\Delta$), the thermal rectification may be related to the geometry of the thermal system due to $k=k(R_i/R_o,L,r,T,T_0, \Delta)$.
For a given temperature range, i.e., $T_0=300\text{K},~|\Delta|=1.0$, we compare the thermal transport phenomena with different radius ratio of the two concentric boundaries: $R_i/R_o=0.2,~0.4,~0.6$.
Numerical simulations are implemented with different $L$ and the results are shown in~\cref{GDT3}.
It can be observed that for a given characteristic length as $R_i/R_o$ increases, the thermal rectification ratio decreases.
The maximum thermal rectification ratio decreases from $31\%$ to $16\%$ as $R_i/R_o$ increases from $0.2$ to $0.6$.
The distributions of the average temperature in the domain $\overline{T}$ are also shown in~\cref{RTTL} and Table.~\ref{TaverageTable}.
It can be observed that for a given characteristic length $L$, as $R_i/R_o$ increases, $\overline{T}_{+}$ increases while $\overline{T}_{-}$ decreases.
In other words, $\overline{\lambda}_{+}$ decreases while $\overline{\lambda}_{-}$ increases so that $\overline{\lambda}_{+}-\overline{\lambda}_{-}$ decreases.
Actually, for a given characteristic length, as $R_i/R_o$ increases, the geometry asymmetry along the radial direction decreases so that the difference between the thermal resistance near the inner boundary and the outer boundary caused by the phonon boundary scattering decreases.
As $R_i/R_o \rightarrow 1.0$, $\overline{\lambda}_{-} \rightarrow \overline{\lambda}_{+}$.
The heat transfer in the concentric ring along the radial direction comes close to that in the cross-plane heat transfer~\cite{MajumdarA93Film}, in which there is no thermal rectification.

\section{Conclusion}
\label{sec:conclusion}
In this study, the radial thermal rectification in the concentric silicon ring from ballistic to diffusive regime is investigated based on the phonon Boltzmann transport equation.
Analytical solutions prove that there is no thermal rectification in the ballistic and diffusive limits.
In the ballistic-diffusive regime, the heat flux prefers to flow from the inner boundary to the outer boundary ($\Delta>0$).
Furthermore, as the characteristic length increases from tens of nanometers to tens of microns, the thermal rectification ratio increases first and then decreases gradually till zero.
Because as $\Delta>0$, the average temperature is lower, which leads to larger phonon mean free path $\overline{\lambda}_{+}$ compared to that as $\Delta<0$ ($\overline{\lambda}_{-}$).
The difference of the average phonon mean free path leads to the stretch or contract of the length-dependent heat flux profiles, especially in the ballistic-diffusive regime.
Different heat flux finally results in thermal rectification.
In addition, the effects of the temperature and radius ratio of the two concentric boundaries are investigated, too.
As the temperature decreases, the maximum thermal rectification ratio decreases.
As the radius ratio of the inner and outer boundaries increases, the thermal rectification ratio decreases for a given characteristic length because the difference between $\overline{\lambda}_{+}$ and $\overline{\lambda}_{-}$ decreases.
The present study offers an idea to realize the thermal rectification in homogeneous materials in the ballistic-diffusive regime by stretching or contracting the length-dependent thermal conductivity or heat flux.

\section*{Conflict of interest}
There is no conflict of interest.

\section*{Acknowledgments}
This work was supported by the National Key Research and Development Plan (No. 2016YFB0600805) and the
National Science Foundation of China (Grants No. 11602091).

\section*{References}
\bibliographystyle{IEEEtr}
\bibliography{phonon}

\end{document}